\documentclass[conference,10pt,table]{IEEEtran}

\usepackage[utf8]{inputenc}
\usepackage{array} 
\usepackage{epsfig}
\usepackage{epstopdf}
\usepackage{times}
\usepackage{framed}
\usepackage{enumitem}
\usepackage{float}
\usepackage{etoolbox}
\usepackage{algpseudocode}
\usepackage{amssymb}
\usepackage{amsmath}
\usepackage{multirow}
\usepackage{url}
\usepackage{lipsum}
\usepackage{soul}
\usepackage{caption}
\usepackage{mathtools}
\usepackage{bbm}
\usepackage{booktabs}
\usepackage{xcolor}
\usepackage{makecell}
\usepackage{pifont}
\usepackage{subfigure}
\usepackage[export]{adjustbox} 
\usepackage{latexsym}
\usepackage{cite}
\usepackage[acronym]{glossaries}
\usepackage{multicol}
\usepackage{eurosym}
\usepackage{xspace}
\usepackage{pdfpages}
\usepackage{verbatim}
\usepackage{stfloats}
\usepackage{mathtools}
\usepackage[norelsize,boxruled]{algorithm2e}

\usepackage{blindtext}

\newcommand{\vv}[1]{\mathbf{#1}}

\newcommand{\Compl}{\mathbb{C}}

\newcommand{\vvhs}[2]{\mathbf{#1}_#2^\mathrm{H}}

\newcommand{\herm}{\mathrm{H}}

\newcommand{\tran}{\mathrm{T}}

\newcommand{\rmG}{\scriptscriptstyle{\mathrm{G}}}
\newcommand{\rmR}{\scriptscriptstyle{\mathrm{R}}}
\newcommand{\rmD}{\scriptscriptstyle{\mathrm{D}}}
\newcommand{\rmT}{\scriptscriptstyle{\mathrm{T}}}

\newcommand{\name}{RISA}

\newacronym{ula}{ULA}{uniform linear array}
\newacronym[plural=BSs, firstplural=base stations (BSs)]{bs}{BS}{base station}
\newacronym{pdf}{pdf}{probability distribution function}
\newacronym{aod}{AoD}{angle of departure}
\newacronym{aoa}{AoA}{angle of arrival}
\newacronym{ue}{UE}{user equipment}
\newacronym{los}{LoS}{line-of-sight}
\newacronym{pla}{PLA}{planar linear array}
\newacronym[plural=RISs, firstplural=reconfigurable intelligent surfaces (RISs)]{ris}{RIS}{reconfigurable intelligent surface}
\newacronym{sdp}{SDP}{semidefinite programming}
\newacronym{sdr}{SDR}{semidefinite relaxation}
\newacronym{sre}{SRE}{smart radio environment}
\newacronym{snr}{SNR}{signal-to-noise ratio}
\newacronym{toa}{ToA}{time-of-arrival}
\newacronym{doa}{DoA}{direction-of-arrival}
\newacronym{mmse}{MMSE}{minimum mean squared error}
\newacronym{peb}{PEB}{position error bound}
\newacronym{oeb}{OEB}{orientation error bound}
\newacronym{rss}{RSS}{received signal strength}
\newacronym{ml}{ML}{machine learning}
\newacronym{rmse}{RMSE}{root-mean-square error}
\newacronym{mmwave}{mm-Wave}{millimeter-wave}
\newacronym{csi}{CSI}{channel state information}
\newacronym{3gpp}{3GPP}{3rd Generation Partnership Project}
\newacronym{sinr}{SINR}{signal-to-interference-plus-noise ratio}
\newacronym{sbr}{SBR}{shooting and bouncing rays}
\newacronym{ura}{URA}{uniform rectangular array}
\newacronym{sncf}{SNCF}{Société nationale des chemins de fer français}
\newacronym{ofdma}{OFDMA}{orthogonal frequency-division multiple access}
\newacronym[plural=CSs, firstplural=candidate sites (CSs)]{cs}{CS}{candidate site}
\newacronym{mrt}{MRT}{maximum ratio transmission}
\newacronym{bca}{BCA}{Block Coordinate Ascent}
\newacronym{fp}{FP}{Fractional Programming}
\newacronym{tdma}{TDMA}{time-division multiple access}
\newacronym{jfi}{JFI}{Jain's fairness index}
\newacronym{rssi}{RSSI}{Receive Signal Strength Indicator}

\newtheorem{problem}{Problem}

\newcommand\Mark[1]{\textsuperscript#1}

\DeclarePairedDelimiter{\norm}{\lVert}{\rVert}

\IEEEoverridecommandlockouts

\begin{document}

\setlength{\textfloatsep}{3pt}

\title{RIS-Aware Indoor Network Planning:\\The Rennes Railway Station Case}

\author{
    \IEEEauthorblockN{Antonio Albanese\Mark{1}\Mark{2}, Guillermo Encinas-Lago\Mark{1},
    Vincenzo Sciancalepore\Mark{1},\\Xavier Costa-P\'erez\Mark{3}\Mark{1}, Dinh-Thuy Phan-Huy\Mark{4},  Stéphane Ros\Mark{5}}
    \IEEEauthorblockA{
	\Mark{1}NEC Laboratories Europe, 69115 Heidelberg, Germany\\
 	\Mark{2}Departamento de Ingeniería Telemática, University Carlos III of Madrid, 28911 Legan\'es, Spain\\
 	\Mark{3}i2cat Foundation and ICREA, 08034 Barcelona, Spain\\
 	\Mark{4} Radio InnOvation (RIO) Department, Orange Innovation (INNOV), 92320 Chatillon, France\\
 	\Mark{5}Société nationale des chemins de fer français, 93210 Saint-Denis, France}
 	\thanks{This work was supported by EU H2020 RISE-6G (grant agreement 101017011) and EU H2020 METAWIRELESS (grant agreement 956256) projects. We thank Guillaume Grao for his support.\newline Email of corresponding author: antonio.albanese@neclab.eu}
}

\maketitle

\begin{abstract}

Future generations of wireless networks will offer unrivalled performance via unprecedented solutions: \emph{metasurfaces} will drive such revolution by enabling control over the surrounding propagation environment, always portrayed as a tamper-proof black box. The \gls{ris} technology, envisioned as the discrete version of a metasurface, can dynamically alter the propagation of the impinging signals by, e.g., steering the corresponding beams towards controllable directions. This will unlock new application opportunities and deliver advanced end-user services.

However, this fascinating solution comes at non-negligible costs: \glspl{ris} require ad-hoc design, deployment and management operations to be fully exploited. In this paper, we tackle the \glspl{ris} placement problem from a theoretical viewpoint, showcasing a large-scale solution on synthetic topologies to improve communication performance while solving the \emph{dead-zone} problem. 
Additionally, our mathematical framework is empirically validated in a realistic indoor scenario, the Rennes railway station, showing how a complex indoor propagation environment can be fully disciplined by an advanced RIS installation.

\end{abstract}

\section{Introduction}
\glsresetall

Recently, the new generation of cellular networks has been successfully integrated and deployed bringing along new business opportunities. However, the revenue-hungry telco operators continuously look for innovative solutions to enable new use cases, which involve new players into the engaged business model. In this context, one emerging technology aims at undermining the classical communication paradigm---that dogmatized the radio propagation environment as an ungovernable box---providing new means to exploit the signal properties: \emph{\glspl{ris}}~\cite{BOL20_ComMag, RIScommag_2021, EuCNC_rise6g,albanese_commag21}.

\begin{figure}[t]
        \center
        \includegraphics[width=.97\linewidth]{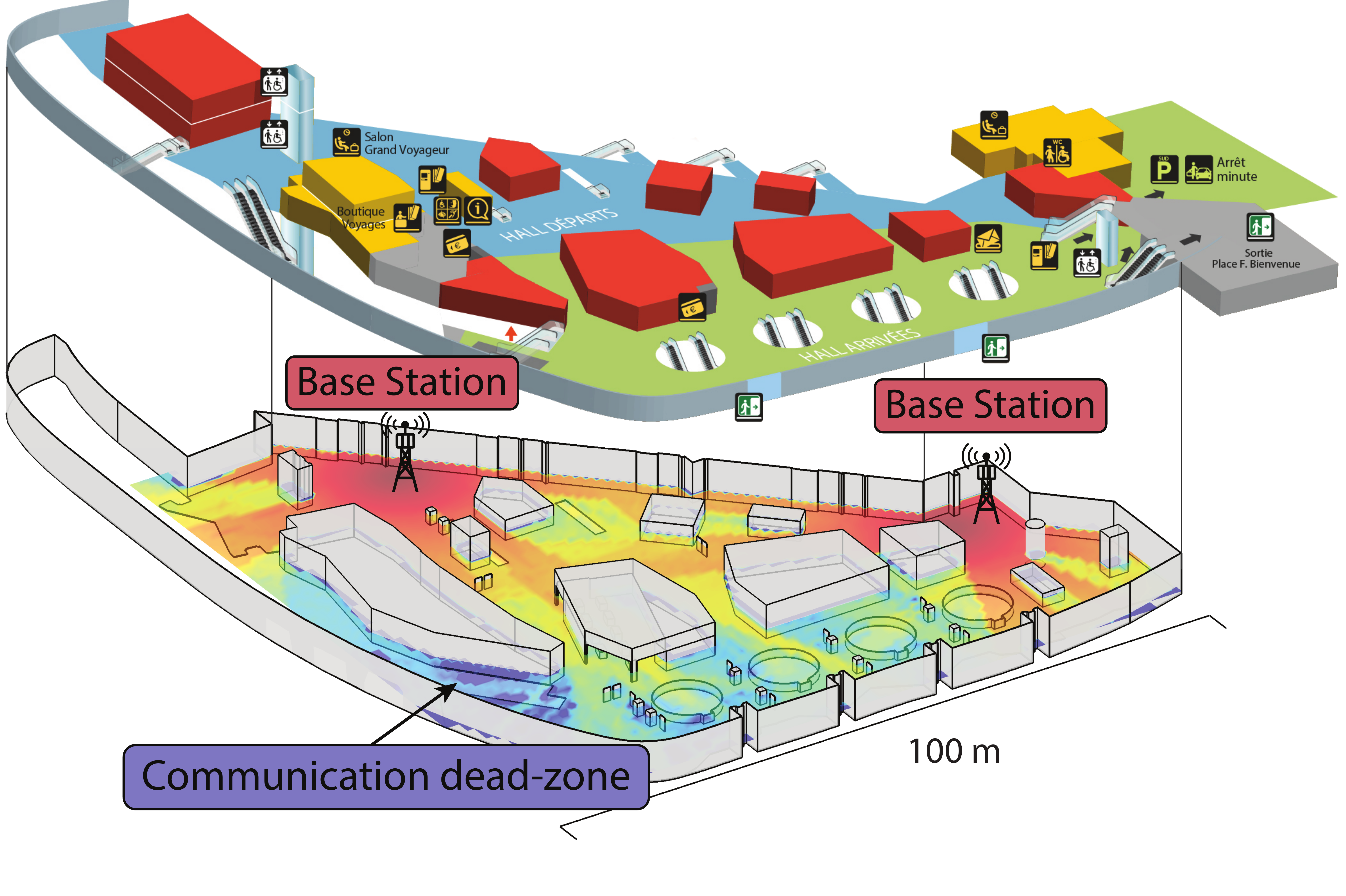}
        \caption{\label{fig:heatmap_no_ris} Railway station topographic map and related power heatmap showing the dead-zone problem (Rennes, France).}
\end{figure}

Agility and flexibility represent the added-value of this solution~\cite{DiRenzo2020,albanese22}: while \glspl{ris} can be dynamically and continuously configured, they draw little power with affordable installation and maintenance costs~\cite{di2019smart}. This makes such a technology the best candidate to solve the mobile dead-zone problem in indoor scenarios by enabling very dense \gls{ris}-based network deployment at low Capital Expenditure (CAPEX). For instance, as shown in Fig.~\ref{fig:heatmap_no_ris}, the existing network infrastructure in a real railway station may fail to guarantee satisfactory performance within the entire environment: \emph{How to solve the dead-zone problem with a very limited investment?}
Ad-hoc \glspl{ris} design and deployment strategies might be the correct answer.
%
Indeed, while \glspl{ris} properly steer the incoming electromagnetic waves towards specific directions, interference is also focused onto unwanted areas, if not properly handled~\cite{metasurface21}. This issue exacerbates the overall deployment complexity calling for advanced optimization techniques to strike the optimal trade-off between \glspl{ris} density and the corresponding spurious detrimental interference.

{\bf Related work.} In the literature, the generic \glspl{bs} deployment problem has been exhaustively investigated, e.g. in~\cite{Andrews2011,Amaldi2003}. The major drawback of such works lies in the isotropic antenna radiation assumption making the problem easy-to-solve via graph-coloring algorithms or convex programming approaches. When dealing with directive transmissions---e.g., millimeter waves (mmWaves) above $6$~GHz---a new degree of freedom is introduced: the beam orientation. Specifically, mmWave \glspl{bs} must be properly placed and electronically oriented to effectively beam towards specific locations leveraging on the available \gls{csi}~\cite{Fascista2019, papir21}. Nonetheless, an optimal \glspl{ris} deployment is even harder to achieve: on the one hand, \glspl{ris} deployment requires prior information on the applied \glspl{ris} configurations; on the other hand, \glspl{ris} configurations can be obtained only upon fixing the \glspl{bs} and \glspl{ris} positions. To overcome this issue and make the analysis tractable, simplistic assumptions on agnostic \glspl{ris} optimization can be done~\cite{Moro2021}.

{\bf Contributions.} Differently, our solution goes one step beyond and jointly tackles the optimal \glspl{ris} placement and configuration problems without any unpractical assumption on the available \gls{csi}. We formulate the overall optimization framework and rely on the well-known \gls{bca}~\cite{Grippo2000} to devise \name{}, a RIS-Aware network planning solution that iteratively derives the \glspl{ris} configurations and optimally places the required number of \glspl{ris} within the area. We $i$) develop a new lightweight ray-tracing model for multi-\gls{ris} scenarios, $ii$) analytically and empirically prove its short convergence time, $iii$) show its efficiency in large-scale scenarios and $iv$) demonstrate outstanding performance in a realistic indoor environment, namely the Rennes Train Station in France, to improve the existing cellular infrastructure of one of the major European operators and solve the dead-zone problem, as shown in Section~\ref{s:performance}. 

\emph{Notation}. We denote matrices and vectors in bold while each of their element is indicated in roman with a subscript. $(\cdot)^{\tran}$ and $(\cdot)^{\herm}$ stand for vector or matrix transposition and Hermitian transposition, respectively. The L$2$-norm of a vector is denoted by $\| \cdot \|$.

\section{System model}
\label{s:system_model}

\addtolength{\topmargin}{+0.28in}
\begin{figure}[t]
        \center
        \includegraphics[width=.95\linewidth]{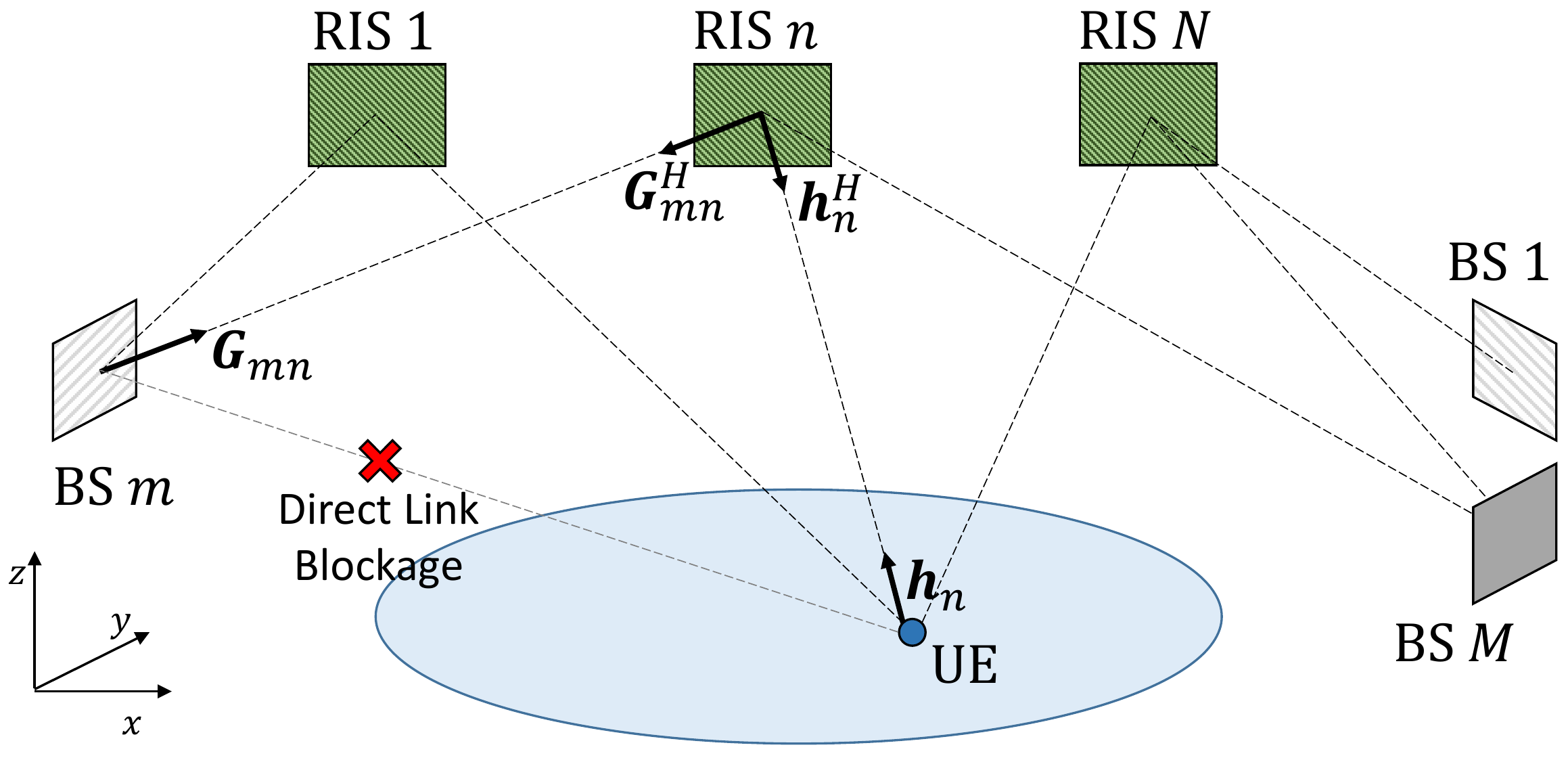}
        \caption{\label{fig:scenario} Geometrical representation of the considered scenario including \glspl{bs}, the \glspl{ris} and one sample \gls{ue}.}
\end{figure}

We consider the 
\gls{ris}-enabled wireless network depicted in Fig.~\ref{fig:scenario}, wherein $N$ \glspl{ris} are deployed to assist $M$ \glspl{bs} to extend their communication coverage in a given area of interest $\mathcal{A}$. We model each \gls{bs} as a \gls{ula} with $N_b$ antennas, and each \gls{ris} as a \gls{pla} with $N_r = N_{h} \times N_v$ reflective elements, where $N_h$ and $N_v$ denote the number of elements in the horizontal plane and the vertical direction of the absolute reference system, respectively. 

We indicate by $\vv{b}_m\in \mathbb{R}^3$, $\vv{r}_n\in \mathbb{R}^3$ and $\vv{u} \in \mathbb{R}^3$ the locations of the $m$-th \gls{bs} center, the $n$-th \gls{ris} center and the typical \gls{ue}, respectively. We assume that the direct \gls{los} links from the \glspl{bs} provide negligible receive power in the target area due to blockage or severe shadowing. Therefore, the communication between \glspl{bs} and \glspl{ue} must be carried out over the reflected link through the \glspl{ris}. In practice, we assume that each \gls{bs} can leverage on multiple \glspl{ris} but each \gls{ris} is used and controlled by a single \gls{bs}, which connects to the on-board \gls{ris} controller via a separate (wired or wireless) reliable control link. Focusing on the downlink transmission, the $m$-th \gls{bs} transmits data to the \gls{ue} over the reflected links through the $n$-th \gls{ris}. Such path can be decomposed into the \gls{los} channel $\vv{h}_n \in \mathbb{C}^{N_r \times 1}$ through which the \gls{ris} reflects the impinging signal towards the \gls{ue}, and the \gls{los} channel $\vv{G}_{mn} \in \mathbb{C}^{N_r\times N_b}$ between the \gls{bs} and the \gls{ris}. 

Let us indicate as $\Lambda_m$, with cardinality $|\Lambda_m|$, the set of \glspl{ris} that are associated with BS $m$. The received downlink signal at the \gls{ue} is given by the superposition of the signals incoming from all \glspl{bs} through their associated \glspl{ris}, namely
\begin{equation}\label{eq:y}
    y \triangleq \sum\limits_{m=0}^{M-1}\sum\limits_{n = 0}^{|\Lambda_m|} \left(\vvhs{h}{n} \vv{\Phi}_n \vv{G}_{mn}\right)  \,  \vv{w}_m s + n \in \Compl,
\end{equation}
where $\vv{\Phi}_n = \mathrm{diag}[\alpha_{n1} e^{j\phi_{n1}}, \dots, \alpha_{nN} e^{j\phi_{nN}}]$ with $\phi_{ni} \in [0, 2\pi]$ and $|\alpha_{ni}|^2 \leq 1$, $\forall i$ indicates the phase shifts and amplitude attenuation introduced by the $n$-th \gls{ris}, $\vv{w}_m \in \Compl^{N_b\times 1}$ is the transmit precoder at the $m$-th BS while $s\in \Compl$ is the transmit signal with $|s|^2=1$, and $n\in \Compl$ is the additive white Gaussian noise term distributed as $\mathcal{CN}(0,\sigma^2)$. 

\begin{figure}[t]
        \center
        \includegraphics[width=.45\linewidth]{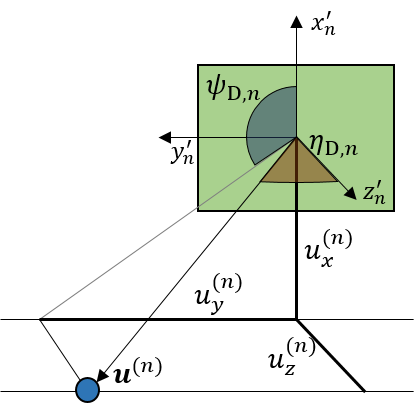}
        \caption{\label{fig:ris_reference} Geometrical representation of one sample user equipment (UE) in the $n$-th RIS reference system.}
\end{figure}

As \gls{3gpp} cellular standards require the \gls{ue} to be served by a single \gls{bs}, we remark that the \gls{ue} receives useful signal only from one \gls{bs}, e.g., the $m$-th \gls{bs}, and suffers from the interference produced by all other \glspl{bs}. 
Therefore, the received \gls{sinr} at the \gls{ue} can be written as
\begin{equation}\label{eq:sinr}
    \mathrm{SINR}(\vv{u}) \triangleq \frac{\left|\sum\limits_{n = 0}^{|\Lambda_m|} \left(\vvhs{h}{n} \vv{\Phi}_n \vv{G}_{mn}\right)  \,  \vv{w}_m\right|^2}{\sum\limits_{\substack{l=0,\\ l \neq m}}^{M-1}\left|\sum\limits_{n = 0}^{|\Lambda_l|} \left(\vvhs{h}{n} \vv{\Phi}_n \vv{G}_{ln}\right)  \,  \vv{w}_l\right|^2 + \sigma^2},
\end{equation}
where the \glspl{bs}-\glspl{ris} and \glspl{ris}-\gls{ue} channels are fully defined by knowing the geometry of the network while the \glspl{ris} configurations and the \glspl{bs} precoders depend on the \glspl{bs}-\glspl{ris} and \glspl{bs}-\gls{ue} associations. 
As shown in Fig.~\ref{fig:ris_reference}, in order to write the channels $\vv{h}_n$ and $\vv{G}_{mn}$, we first consider $N$ reference systems with origin in
the center of each \gls{ris} and the $(x',y')$-plane
lying on the RIS surface.
Hence, the coordinates of the \gls{ue} in the reference system of the $n$-th \gls{ris} can be obtained as 
    $\vv{u}^{(n)} = \vv{R}_n \vv{u}$,
where 
\begin{equation}
    \vv{R}_n \triangleq 
    \begin{pmatrix} 
    \hat{\vv{r}}_{n,x'} & \hat{\vv{r}}_{n,y'} & \hat{\vv{r}}_{n,z'}
    \end{pmatrix}     \in \mathbb{R}^{3 \times 3},
\end{equation}
with $\hat{\vv{r}}_{n,x'}$, $\hat{\vv{r}}_{n,y'}$ and $\hat{\vv{r}}_{n,z'} \in \mathbb{R}^3$ representing the coordinates of the $n$-th \gls{ris} reference system axes in the absolute reference system. Furthermore, we denote by $\psi_{\rmD,n}$ and $\eta_{\rmD,n}$ the azimuth and the zenith \gls{aod} for the communication link from the \gls{ris} to the \gls{ue}. Therefore, the \gls{ris} array response vector is given by
\begin{align}
    \vv{b}_{\rmT,n}(\vv{u}) \triangleq \, &  \vv{b}_y (\psi_{\rmD,n}, \eta_{\rmD,n}) \otimes \vv{b}_z (\psi_{\rmD,n}, \eta_{\rmD,n}) \in \Compl^{N_r\times 1}\label{eq:pla}\\
    = \,&[1, e^{j2\pi\delta\sin(\psi_{\rmD,n})\sin(\eta_{\rmD,n})}, \dots, \nonumber\\
    &e^{j2\pi\delta(N_y -1)\sin(\psi_{\rmD,n})\sin(\eta_{\rmD,n})}]^\mathrm{T} \nonumber \\
    &\otimes[1, e^{j2\pi\delta\cos(\psi_{\rmD,n})\sin(\eta_{\rmD,n})}, \dots, \nonumber \\
    &e^{j2\pi\delta(N_x -1)\cos(\psi_{\rmD,n})\sin(\eta_{\rmD,n})}]^\mathrm{T} ,
\end{align}
where $\delta$ indicates the antenna spacing-wavelength ratio. We refer to $\Omega_{\rmT,n}(\vv{u}) \triangleq \cos(\psi_{\rmD,n})\sin(\eta_{\rmD,n}) = \frac{u_x^{(n)}}{\norm{\vv{u}^{(n)}}}$ and $\Psi_{\rmT,n}(\vv{u}) \triangleq \sin(\psi_{\rmD,n})\sin(\eta_{\rmD,n}) = \frac{u_y^{(n)}}{\norm{\vv{u}^{(n)}}}$ as the spatial frequencies along the $x'_n$ and the $y'_n$-axis corresponding to the \gls{aod} towards the \gls{ue} at absolute coordinates $\vv{u}$. Therefore, the \gls{los} $n$-th \gls{ris}-\gls{ue} channel is given by 
\begin{equation}
    \vv{h}_n(\vv{u}) \triangleq \sqrt{\gamma_n(\vv{u})} \, \vv{b}_{\rmT.n}(\vv{u}) \in \Compl^{N_r \times 1}, \label{eq:h_n}
\end{equation}
where $\gamma_n(\vv{u}) \triangleq d_n(\vv{u})^{-\beta}$ is the channel power gain with $d_n(\vv{u}) = \norm{\vv{r}_n - \vv{u}}$ being the Eucledian distance between the \gls{ris} and the \gls{ue}. In a similar way, 
the \gls{los} channel between the $m$-th \gls{bs} and the $n$-th \gls{ris} can be written as 
\begin{equation}
    \vv{G}_{mn} \triangleq \sqrt{\gamma_{\rmG_{mn}}} \, \vv{b}_{\rmR,n}(\vv{b}_m) \vvhs{a}{m}(\vv{r}_n) \in \Compl^{N_r\times N_b},
    \label{eq:G_mn}
\end{equation}
where $\gamma_{\rmG_{mn}} \triangleq d_{mn}^{-\beta}$ is the channel power gain with $d_{mn} = \norm{\vv{b_m}-\vv{r}_n}$, $\vv{a}_{\rmR,n}(\vv{b}_m)$ is the array response vector at the \gls{ris} corresponding to the \gls{aoa} from \gls{bs} m, which is derived analogously to Eq.~\eqref{eq:pla}, and $\vv{a}_m(\vv{r}_n)$ indicates the \gls{bs} array response, defined as 
\begin{equation}
    \vv{a}_m(\vv{r}_n) \triangleq [1, \dots, e^{j2\pi\delta(M-1)\cos(\theta_{\rmD,mn})}]^\mathrm{T} \in \Compl^{N_b \times 1},
\end{equation}
where $\theta_{\rmD,mn}$ represents the \gls{aod} from the $m$-th \gls{bs} to the $n$-th \gls{ris}.


\section{Problem Formulation}
\label{s:problem}

\textbf{Analytical tractability.} The solution to our multi-\gls{ris} planning problem requires determining the optimal \glspl{ris} deployment to provide coverage within the target area, e.g., by maximizing the worst-case received \gls{sinr} at all locations $\vv{u}$. 
To this aim, we need to jointly optimize the active transmit beamformers at the \glspl{bs} as well as the \glspl{ris} placement, their passive beamforming configurations, and their controlling \glspl{bs}, which in turn dictate the optimal end-to-end \gls{bs}-\gls{ue} associations. The resulting optimization problem is highly non-convex and extremely difficult to tackle due to the intricate coupling between the \glspl{bs}-\glspl{ris} and \glspl{bs}-\gls{ue} associations, and the joint active-passive beamforming configurations throughout the network. For instance, even for a given \gls{bs}-\gls{ris}-\gls{ue} association, jointly optimizing the beamforming at the \gls{bs} and the \gls{ris} does not yield a closed-form formulation but rather requires tackling a non-convex problem by alternatively solving the two separate beamforming optimizations until convergence~\cite{Wu2019}.  

Therefore, for the sake of analytical tractability, we consider propagation paths involving only first-order \gls{ris} reflections, and assume a cellular-like architecture in which each \gls{ris} provides coverage to one contiguous subarea, thus reducing the scope of the interference generated by the remaining \glspl{ris} to the sheer overlapping area edges. 
We would like to highlight that at planning stage, the \glspl{ris} beamforming design for area coverage enhancement cannot take advantage of the knowledge of the instantaneous \gls{csi} of any particular \gls{ue} in the area. Hence, although \glspl{ris} controlled by the same \gls{bs} can be configured to cover the same subarea, it is highly complex to enforce in-phase constructive interference of signals incoming from different \glspl{ris} even if transmitted by the same \gls{bs}. 

Let us consider the \gls{ue} to be inside the subarea served by BS $m$ through RIS $n$. In these conditions, its received \gls{sinr} can be then approximated by its \gls{snr}, which is defined as 
\begin{equation}\label{eq:snr}
    \mathrm{SNR}(\vv{\Phi}_n, \vv{w}_m, \vv{u}) \triangleq \frac{\left| \vvhs{h}{n} \vv{\Phi}_n \vv{G}_{mn}  \,  \vv{w}_m\right|^2}{\sigma^2},
\end{equation}
where $\vv{\Phi}_n$ and $\vv{w}_m$ need to be optimized. 

\textbf{Optimization variables.}
We assume that the \glspl{ris} are deployed only at specific locations, i.e. \glspl{cs}, to reflect the fact that network operators are required to meet logistical, administrative and physical constraints in real-life scenarios. Nonetheless, in the absence of \glspl{cs}, our multi-\gls{ris} planning may be likewise executed by considering any sampling of the deployment area. For the sake of simplicity, we assume that the \glspl{cs} set matches the set $\{\vv{r}_n\}_{n=1}^{N}$, namely the candidate \glspl{ris} positions are pre-defined and we aim at identifying where to actually deploy \glspl{ris} among them. Besides, we sample the target area by means of $T$ test points $\vv{u}_t \in \mathcal{A}$, wherein we optimize the \gls{snr} of the typical \gls{ue}\footnote{Ideally, the test point distribution should match the expected distribution of the users in the target area but the problem formulation remains valid for any distribution of users.}. 
Our planning solution outputs the set of \glspl{ris} to be deployed while providing the optimal \gls{bs}-\gls{ris}-\gls{ue} association at each est point. We thus introduce decision variables $\vv{x} \in \{0,1\}^{N}$ and $\vv{y}\in \{0,1\}^{T \times M \times N}$, whose elements $x_n$ and $y_{tmn}$ indicate whether a \gls{ris} is deployed at CS $n$, and the association between the typical UE at test point $\vv{u}_t$, \gls{bs} m and \gls{ris} at \gls{cs} $n$, respectively. 

\textbf{\glspl{ris} planning.} We can now formulate the multi-\gls{ris} coverage enhancement problem as the following
\begin{subequations}
\begin{problem}[Multi-\gls{ris} coverage enhancement]\label{problem:max_snr}
\begin{align}
   \displaystyle \max_{\vv{\Phi}_n, \vv{w}_m, \vv{x}, \vv{y}} & \displaystyle \min_{\vv{u}_t} \sum\limits_{m,n} y_{tmn}\left| \vvhs{h}{n} \vv{\Phi}_n \vv{G}_{mn}  \,  \vv{w}_m\right|^2 \label{eq:max_min_snr}\\
   \displaystyle \textup{s.t.} & \ \ \displaystyle |\Phi_{n,ii}|^2 \leq  1, \hspace{2.43cm} \forall n, \forall i, \label{eq:phi_con} \\
   & \ \ \displaystyle\norm{\vv{w}_m}^2 \leq P, \hspace{2.4cm} \forall m, \label{eq:precoder_con} \\
    & \ \ \displaystyle  y_{tmn} \,\vv{\hat{r}}_{n,x'}^{\tran} (\vv{u}_t - \vv{r}_n) \geq 0,  \hspace{0.65cm} \forall t, \forall m, \forall n, \label{eq:test_point_orientation_con}\\
    & \ \ \displaystyle y_{tmn} \, \vv{\hat{r}}_{n,x'}^{\tran} (\vv{b}_m - \vv{r}_n) \geq 0,  \hspace{0.53cm} \forall t, \forall m , \forall n, \label{eq:base_station_orientation_con}\\
    & \ \ \displaystyle y_{tmn} \leq x_n, \hspace{2.63cm} \forall t, \forall m, \forall n, \label{eq:deployment_con}\\
    & \ \ \displaystyle \sum\limits_{m,n} y_{tmn} = 1, \hspace{2.22cm} \forall t, \label{eq:coverage_con} \\ 
    & \ \ \displaystyle \sum\limits_m \max\limits_t y_{tmn} \leq 1, \label{eq:one_bs_con} \hspace{1.5cm} \forall n\\
    & \ \ \displaystyle \sum\limits_{n} x_n = L, \label{eq:budget_con}\\
    & \ \ \displaystyle x_n \in [0,1], \quad y_{tmn} \in [0,1], \hspace{0.4cm} \forall t, \forall m, \forall n \label{eq:binary_con},
\end{align}
\end{problem}
\end{subequations}
where we omit the constant noise term $\sigma^2$ and refer to the available transmit power at the \glspl{bs} as $P$. The constraint in Eq.~\eqref{eq:phi_con} ensures that the \glspl{ris} are passive while the one in Eq.~\eqref{eq:precoder_con} enforces that the \glspl{bs} power budget is satisfied by each precoder $\vv{w}_m$. Constraints \eqref{eq:test_point_orientation_con} and \eqref{eq:base_station_orientation_con} guarantee that each \gls{ris} can respectively serve a test point or be assigned to a \gls{bs} only if they front the \gls{ris}, i.e. only if the vector originated in the \gls{ris} and pointing towards the test point or the \gls{bs} has a positive projection on the \gls{ris} orientation vector $\vv{\hat{r}}_{n,x'}$. Moreover, constraint~\eqref{eq:deployment_con} states that a \gls{ris} should be deployed only if at least the \gls{ue} located at one test point would exploit it, whereas constraint~\eqref{eq:coverage_con} reflects the fact that each test point must be covered by only one \gls{ris}. Constraint~\eqref{eq:one_bs_con} forces each \gls{cs} to be associated to at most one \gls{bs} and, lastly, we enforce the number of deployed \glspl{ris} to be equal to $L$ in constraint~\eqref{eq:budget_con}, where $L$ is the number of \glspl{ris} to be deployed by the network operator.

\section{RIS-aware Network Planning}
\label{s:planning}

Even disregarding the interference, Problem~\eqref{problem:max_snr} is still highly complex due to its objective function in Eq.~\eqref{eq:max_min_snr} being the sum of non-convex elements, and the binary constraints in Eq.~\eqref{eq:binary_con} that make it combinatorial. Moreover, as already mentioned in Section~\ref{s:problem}, the lack of knowledge about the instantaneous \glspl{ue} \gls{csi} in the target area during a realistic access procedure invalidates the option of jointly configuring the \glspl{ris} and \glspl{bs} beamformers per \gls{ue}~\cite{Mursia2021}. Therefore, we decouple the \glspl{ris} and \glspl{bs} beamforming configurations from the planning problem itself by configuring each \gls{ris} to provide coverage to one contiguous subarea and assuming that each \gls{bs} radiates all its available power towards each of its associated \glspl{ris} in a \gls{tdma} fashion. 
In other words, we assume that the \glspl{ris} have a single-beam radiation pattern and that they do not serve more than one subareas, thereby guaranteeing that all locations belonging to one subarea are served by a single \gls{bs} through one single \gls{ris}. Given sufficient coverage in the area, multiple users in each subarea can be separated by conventional multiple access techniques, such as \gls{tdma} or \gls{ofdma}.     


It can be easily observed from Eq.~\eqref{eq:snr} that the \gls{snr} at the \gls{ue} $\vv{u}_t$ provided by \gls{bs} $m$ through \gls{ris} $n$ can be equivalently written as  
    $\mathrm{SNR}(\vv{\Phi}_n, \vv{w}_m, \vv{u}_t) = \frac{g_1(\vv{\Phi}_m, \vv{w}_m, \vv{u}_t)}{g_2(\vv{u}_t)}$,
where $g_1(\vv{\Phi}_n, \vv{w}_m, \vv{u}_t)$ provides the overall array gain due to the cascaded active and passive beamformings, while $g_2(\vv{u}_t)$ accounts for the concatenated \gls{bs}-\gls{ris}-\gls{ue} pathloss.    
Following~\cite{Lu2021}, the \gls{ris} configuration can be obtained by means of 3D beam broadening and flattening, namely by partitioning the \gls{ris} into multiple sub-arrays of smaller size and optimizing their phase shifts to shape one single flattened beam whose beamwidth can be properly tuned to match the size of the target subarea. 
In particular, by denoting the subarea covered by \gls{ris} $n$ by $\mathcal{A}_n$ and assuming $\vv{u}_t \in \mathcal{A}_n$, the resulting \gls{bs}-\gls{ris} gain can be written as 
\begin{equation}
    g_1(\vv{\bar{\Phi}}_n, \bar{\vv{w}}_m, \vv{u}_t) \approx \frac{N_h^2}{\Delta_{n,x'}N_h \delta} \frac{N_v^2}{\Delta_{n,y'}N_v \delta}, \quad \forall \vv{u}_t \in \mathcal{A}_n,
\end{equation}
where $\bar{\vv{\Phi}}_n$ is derived by means of beam broadening and flattening, and $\vv{w}_m$ is the \gls{mrt} precoder, which depends only on $\vv{G}_{mn}$. Besides, $\Delta_{n,x'}$ and $\Delta_{n,y'}$ respectively denote the desired spans of the spatial frequency
deviations along the horizontal $x'_n$ and vertical $y'_n$-axis of \gls{ris} $n$ to cover its subarea and are defined as 
\begin{align}
    \Delta_{n,x'} \triangleq \max_{\vv{u}_t \in \mathcal{A}_n} \Omega_{\rmT,n} (\vv{u}_t) - \min_{\vv{u}_t \in \mathcal{A}_n} \Omega_{\rmT,n} (\vv{u}_t),\\
    \Delta_{n,y'} \triangleq \max_{\vv{u}_t \in \mathcal{A}_n} \Psi_{\rmT,n} (\vv{u}_t) - \min_{\vv{u}_t \in \mathcal{A}_n} \Psi_{\rmT,n} (\vv{u}_t).
\end{align}
The overall pathloss experienced by the \gls{ue} at coordinates $\vv{u}_t$ is given by $g_2(\vv{u}_t) = d^{\beta}_{mn} d^{\beta}_n(\vv{u}_t)$.
Therefore, we can state the following equivalent formulation for Problem~\eqref{problem:max_snr}, i.e.
\begin{subequations}
\begin{problem}[Multi-\gls{ris} planning]\label{problem:max_snr_eq}
\begin{align}
   \displaystyle \max_{\vv{x}, \vv{y}, \vv{\Delta}_{x'}, \vv{\Delta}_{y'}}  & \displaystyle \min_{\vv{u}_t} \sum\limits_{m,n} y_{tmn} \frac{1}{\Delta_{n,x'}\Delta_{n,y'}} \frac{1}{d^{\beta}_{mn} d^{\beta}_n(\vv{u}_t)} \label{eq:max_min_snr_eq}\\
   \displaystyle \textup{s.t.} & \ \ \sum\limits_m y_{tmn}\left| \Omega_{\rmT,n}(\vv{u}_t) - \Omega_{\rmT,n} (\vv{u}_k)\right| \leq \Delta_{n,x'}, \nonumber\\
    & \ \ \hspace{3.3cm} \forall n, \forall \vv{u}_t,\vv{u}_k \in \mathcal{A}, \label{eq:omega_con}\\
    & \ \ \sum\limits_m y_{tmn}\left| \Psi_{\rmT,n}(\vv{u}_t) - \Psi_{\rmT,n} (\vv{u}_k)\right|  \leq \Delta_{n,y'}, \nonumber\\
    & \ \ \hspace{3.3cm} \forall n, \forall \vv{u}_t,\vv{u}_k \in \mathcal{A}, \label{eq:psi_con}\\
    & \ \ \Delta_{n,x'} \geq \frac{1}{N_h \delta}, \quad \Delta_{n,y'} \geq \frac{1}{N_v \delta}, \hspace{0.65cm} \forall n, \label{eq:deltas_min_con} \\
    & \ \  \eqref{eq:test_point_orientation_con}, \eqref{eq:base_station_orientation_con}, \eqref{eq:deployment_con}, \eqref{eq:coverage_con}, \eqref{eq:one_bs_con}, \eqref{eq:budget_con}, \eqref{eq:binary_con},\nonumber
\end{align}
\end{problem}
\end{subequations}
in which we define $\vv{\Delta}_{x'} \triangleq [\Delta_{1,x'}, \dots, \Delta_{N,x'}]$, $\vv{\Delta}_{y'} \triangleq [\Delta_{1,y'}, \dots, \Delta_{N,y'}]$ while we omit the constant terms. In this equivalent formulation, we introduce the constraints in Eqs.~\eqref{eq:omega_con} and~\eqref{eq:psi_con} in order to guarantee that each test point served by \gls{ris} $n$ lies within the coverage determined by its spatial frequency span. In Eq.~\eqref{eq:deltas_min_con}, we enforce that the spatial frequency spans $\vv{\Delta}_{x'}$ and $\vv{\Delta}_{y'}$ are at least as wide as the minimum beamwidth obtained by considering a single sub-array while performing the \glspl{ris} configuration via beam broadening and flattening, as by~\cite{Lu2021}.   


\section{Large-scale planning algorithm}
\label{s:algorithm}
Hereafter, we design our multi-\gls{ris} planning algorithm, i.e., \name{}. Let us first consider a continuous relaxation of Problem~\eqref{problem:max_snr_eq} by letting $\vv{x} \in [0,1]^N$ and $\vv{y} \in [0,1]^{T \times M \times N}$ in constraint~\eqref{eq:binary_con}. We can tackle such problem by means of \gls{bca}, namely by iteratively solving the problem for one block of optimization variables while keeping all the others fixed. Notably, although non-convex in general, the continuous relaxation of the problem is jointly convex in the block of variables $\vv{x},\vv{y}$ while it is still non-convex neither in $\vv{\Delta}_{x'}$, nor in $\vv{\Delta}_{y'}$, as the respective objective functions are convex and their maximization leads to a non-convex problem \textit{per se}. 
In order to solve the problem for $\vv{\Delta}_{x'}$ (or, similarly, for $\vv{\Delta}_{y'}$), we can rearrange Eq.~\eqref{eq:max_min_snr_eq} as 
\begin{align}
   \displaystyle \max_{\vv{\Delta}_{x'}}  \quad \displaystyle \min_{\vv{u}_t} \sum\limits_{m,n} \frac{y_{tmn}}{d^2_{mn} d^2_n(\vv{u}_t)\Delta_{n,y'}} \frac{1}{\Delta_{n,x'}},  \label{eq:max_min_snr_eq_deltax}
\end{align}
and observe that the resulting subproblem belongs to the \gls{fp} umbrella, being Eq.~\eqref{eq:max_min_snr_eq_deltax} a \textit{sum of functions of ratios}. Therefore, we can leverage on the Quadratic Transform~\cite{Shen2018} and write it equivalently as 
\begin{align}
   \displaystyle \max_{\vv{z}_{x'},\vv{\Delta}_{x'}}  \, \displaystyle \min_{\vv{u}_t} \sum\limits_{m,n} \frac{y_{tmn}}{d^2_{mn} d^2_n(\vv{u}_t)\Delta_{n,y'}} (2z_{n,x'} \!-\! z^2_{n,x'}\Delta_{n,x'}),  \label{eq:max_min_snr_eq_deltax_transform}
\end{align}
where $\vv{z}_{x'} \in \mathbb{R}^N$ is an auxiliary optimization variable. The resulting subproblem is now convex in $\vv{z}_{x'}$ and in $\vv{\Delta}_{x'}$ separately, thus likewise solvable by means of a nested \gls{bca}. 

Therefore, the solution of the continuous relaxation of Problem~\eqref{problem:max_snr_eq} consists of a double-nested \gls{bca} whose outer loop iteratively considers the three blocks of variables $\vv{x}$ and $\vv{y}$, $\vv{\Delta}_{x'}$, $\vv{\Delta}_{y'}$, while its inner loops solve the subproblems in $\vv{\Delta}_{x'}$ and $\vv{\Delta}_{y'}$ by introducing auxiliary variables $\vv{z}_{x'} \in \mathbb{R}^N$ and $\vv{z}_{y'} \in \mathbb{R}^N$, respectively. We would like to highlight that, by dealing with a convex problem at each stage, the double-nested \gls{bca} is guaranteed to converge to a stationary point~\cite{Grippo2000}. 

\textbf{Binary solution.} The binary deployment variable $\vv{x}^*$ are recovered by rounding the highest $L$ elements of $\vv{x}$ to $1$ while setting the other $N-L$ to $0$. Next, we establish the binary associations $\vv{y}^*$ by considering only activated \glspl{cs} $\vv{b}_n$ such that $x^*_n = 1$. In particular, we iteratively associate each test point $\vv{u}_t$ by setting to $1$ the highest element $\vv{y}$ among the ones corresponding to the activated \glspl{cs}. Concurrently, we update the values of $\Delta_{n,x'}$ and $\Delta_{n,y'}$ to the minimum spatial frequency spans satisfying the constraints in Eqs.~\eqref{eq:omega_con},~\eqref{eq:psi_con}. 

We depict the overall high-level algorithm in Algorithm~\ref{alg:deployment}.

\begin{algorithm}[t]
    \DontPrintSemicolon
    Initialize $\Delta_{n,x'} = \Delta_{n,y'} = 2$, $n = 1, \dots, N$ \;
    \Repeat(\hfill\emph{Outer BCA loop}){convergence of objective function Eq.~\eqref{eq:max_min_snr_eq} in Problem~\ref{problem:max_snr_eq}}{
        Solve the continuous relaxation of Problem~\ref{problem:max_snr_eq} jointly for $\vv{x}$ and $\vv{y}$ \;
        \Repeat(\hfill\emph{First inner BCA loop }){convergence of objective function in Eq.~\eqref{eq:max_min_snr_eq_deltax}}{
            Solve the transformed problem in $\vv{\Delta}_{x'}$ for $\vv{z}_{x'}$\;
            Solve the transformed problem in $\vv{\Delta}_{x'}$ for $\vv{\Delta}_{x'}$\;   
        }
        \Repeat(\hfill\emph{Second inner BCA loop }){convergence of objective function in Eq.~\eqref{eq:max_min_snr_eq_deltax} for $\Delta_{y'}$}{
            Solve the transformed problem in $\vv{\Delta}_{y'}$ for $\vv{z}_{y'}$\;
            Solve the transformed problem in $\vv{\Delta}_{y'}$ for $\vv{\Delta}_{y'}$\;   
        }
      }
    Round $\vv{x}$ and $\vv{y}$ to derive binary $\vv{x}^*$ and $\vv{y}^*$ as by Section~\ref{s:algorithm}\;
\caption{\gls{ris}-Aware network planning (\name{})}
\label{alg:deployment}
\end{algorithm}

\section{Performance evaluation}
\label{s:performance}

We first evaluate \name{} via Monte Carlo simulations considering synthetic network topologies. Subsequently, we benchmark \name{} on the real network topology installed in the \textit{Rennes railway station}, France, provided by the European network operator \textit{Orange}, wherein \glspl{ris} Candidate Sites (CSs) are properly handpicked on the station floor plan and realistic \gls{snr} values are obtained via ray tracing. Simulation parameters based on realistic values are listed in Table~\ref{tab:parameters}, unless otherwise stated. 

\begin{table}[t]
\caption{Simulation parameters.}
\label{tab:parameters}
\centering
\resizebox{\linewidth}{!}{%
\begin{tabular}{cc|cc|cc}
\textbf{Parameter} & \textbf{Value} & \textbf{Parameter} & \textbf{Value} & \textbf{Parameter} & \textbf{Value}\\  
\hline
\rowcolor[HTML]{EFEFEF}
$P$ & $28$ dBm  & $f$  & $26$ GHz & $\sigma^2$  & $-80$ dBm     \\
$\beta$ & 2 &  $\mu$ & $0.5$   & A  &     $100 \,\text{m} \times 100 \,\text{m} $\\
\rowcolor[HTML]{EFEFEF}
BSs ($M$) & 2 & CSs ($N$) &\{10,20,30\} &$T$ & $100$   \\
$N_b$ & 2 & $N_r$ & $350 \times 175$ & $N_{ref}$ & 2 \\
\hline
\end{tabular}%
}
\end{table}

\begin{figure}[t!]
    \centering  
    \subfigure[\name{} \gls{snr} performance.]
    {
        \includegraphics[clip,width=.46\linewidth ]{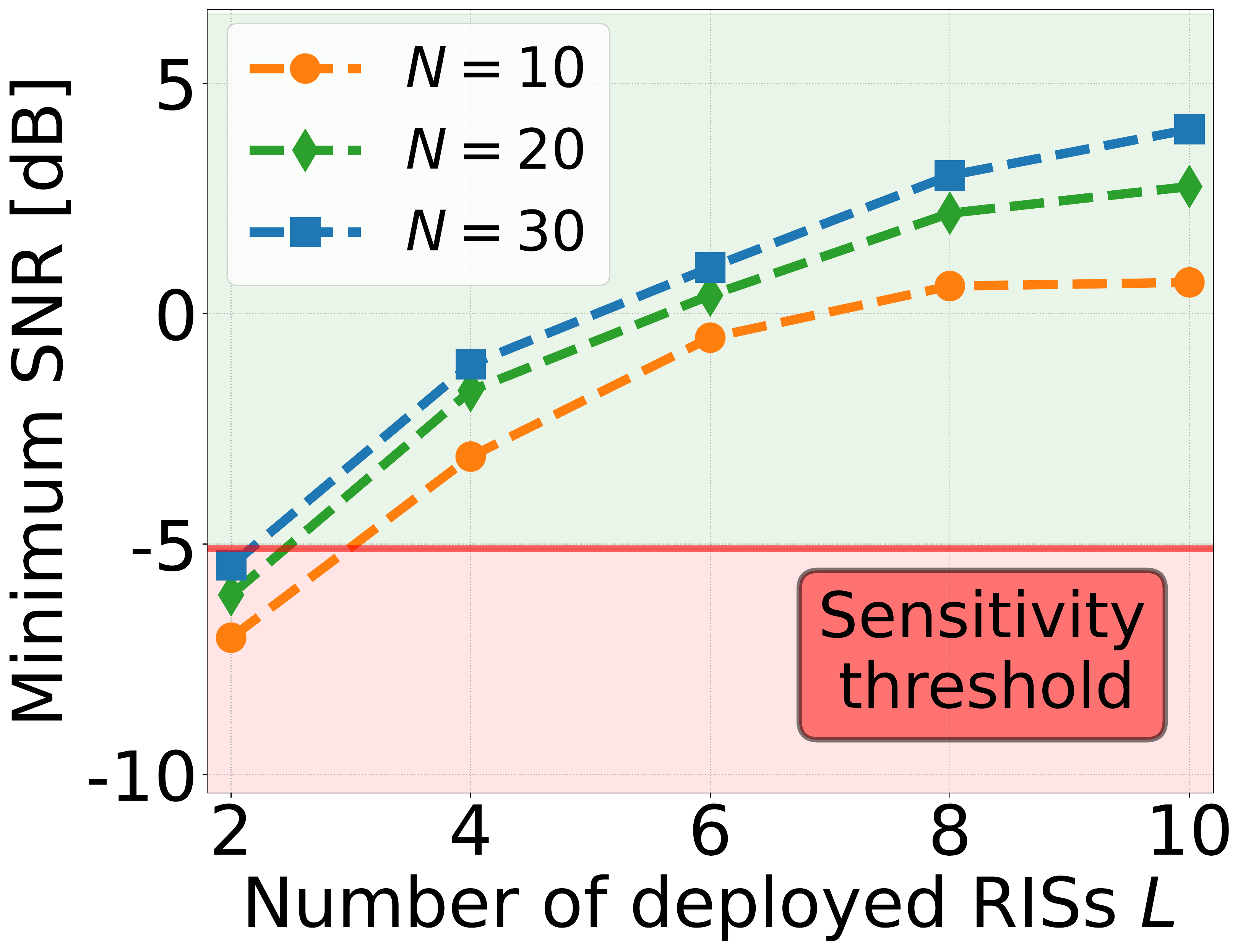}
        \label{fig:snr_vs_nris}
    }
    \subfigure[\name{} fairness performance.]
    {
        \includegraphics[clip,width=.46\linewidth]{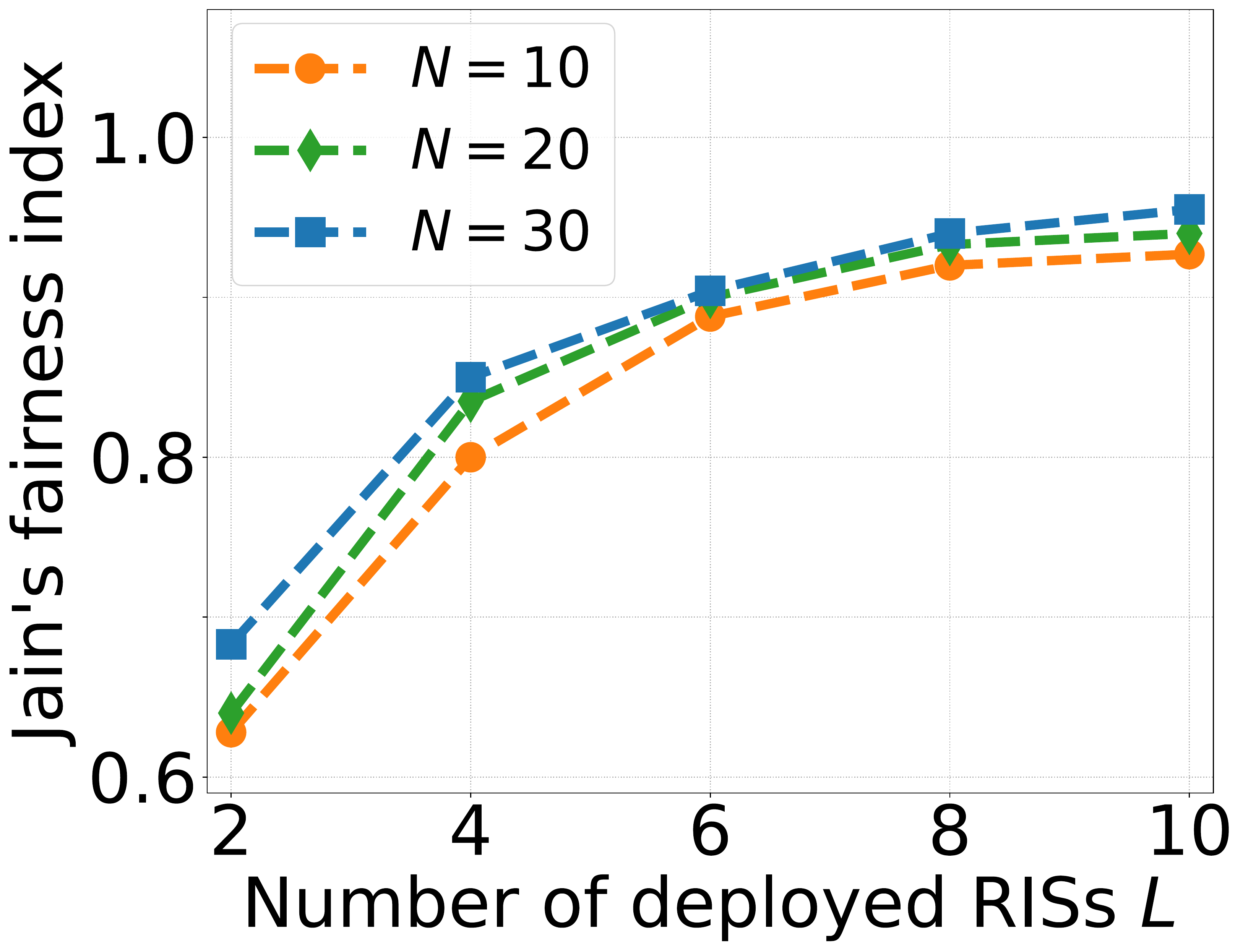}
        \label{fig:jfi_vs_nris}
    }
    \vspace{-1mm}
    \caption{\label{fig:perf_vs_nris} \name{} performance with different numbers of deployed \glspl{ris} and available Candidate Sites (CSs) via Monte Carlo simulations considering synthetic topologies.}
\end{figure}

\textbf{Synthetic topologies.} We consider the target area to be a square surface with area $A = 100\, \text{m} \times 100\, \text{m}$. Besides, we assume that $M = 2$ \glspl{bs} are placed at the bottom-left and upper-right corners of the area, namely $\vv{b}_1 = [0,0,5.5]^\tran$, $\vv{b}_2 = [10^3,10^3,5.5]^\tran$, while we evaluate the \gls{snr} performance at $T = 100$ test points uniformly distributed in the target area on the plane $z = 1.5$. We average the results over $10^3$ Monte Carlo executions. In Fig.~\ref{fig:snr_vs_nris}, we show the performance of \name{} in terms of minimum \gls{snr} experienced in the target area with respect to the number of deployed \glspl{ris} $L$ for different numbers of available \glspl{cs} $N = \{10,20,30\}$ on the plane $z = 5.5$. The horizontal line indicates the minimum \gls{snr} threshold to meet the receiver sensitivity. As expected, the minimum \gls{snr} shows a positive monotonic behavior with decreasing relative increments, thus suggesting the existence of an optimal value for $L$, e.g. $L = 8$ deployed \glspl{ris} for $N = 10$ candidate sites. Besides, increasing the number of \glspl{cs} does not significantly benefit the overall performance, provided that the number of \glspl{cs} is big enough to obtain a good sampling of the target area (on average). The \gls{snr} fairness among test points, measured by means of the \gls{jfi}~\cite{Sediq2013}, shows a similar behavior\footnote{Note that, to obtain meaningful numerical results for the \gls{jfi}, we prevent the received power from exceeding a given maximum, i.e. $-65$ dBm, which provides excellent \gls{rssi}.} in Fig.~\ref{fig:jfi_vs_nris}, validating our \textit{max-min} objective function design choice to enhance coverage in the whole area.

\textbf{Rennes station.} We execute the ray-tracing simulation in MATLAB R2021b using a simplified 3D model of the main floor of the Rennes railway station in France. 
The scenario follows the most prominent obstacles and elements filling the volume object of study. 
Highly convoluted or unknown elements (e.g., the ceiling, composed of many structural, functional, and decorative beams, as well as the tubing, etc.) have been left as holes to simulate the lack of significant, predictable reflections and lessen the computational burden. The resulting model has $579$ triangles, $1582$ edges and $1053$ vertexes and is depicted in Fig.~\ref{fig:heatmap_no_ris}.
We simulate the \glspl{bs} with $N_b = 2$ and transmit power $P = 28$ dBm at $f = 26$ GHz, as in the real network deployment by \textit{Orange}. Besides, we implement \gls{sbr} in order to derive the possible paths to reach any given test point~\cite{Brem2015}. We linearly combine the power received at any test point from different paths assuming a (uniformly distributed) random phase for each individual path at the \gls{ue} side, thereby accounting for random external factors (e.g., thermal expansion) that could alter the path lengths by a non-negligible fraction of a wavelength $\lambda = 1/f$, given the large ratio between the station distances and the wavelength~\cite{rappaport_2001}. The maximum number of reflections is set to $N_{ref} = 2$ as higher-order reflections provide little contribution to the received power. 

\textbf{\gls{ris} ray-tracing model.} We would like to underline that \glspl{ris} are novel network devices, thereby not yet widely implemented in conventional ray-tracers. Therefore, we devise a new lightweight technique to compute the impinging power on the \gls{ris} surface, the \gls{ris} power reflection and the \gls{ris} beampattern. To estimate the impinging power on the \gls{ris} surface, we assess the power received at the \gls{ris} center by one of its elements modeled as a cosine antenna (with exponent parameter $\mu = 0.5$) 
and multiply this value by the number of \gls{ris} elements $N_r$.  
Hence, we simulate the \gls{ris} controlled reflections by considering outgoing rays originated on the \gls{ris} surface with power equal to the \gls{ris} impinging power. The \gls{ris} emissive beampattern is modeled as a \gls{ura} of $N_r$ cosine antennas, where the phase shifts of each element is controlled with narrow-band phase-shift beamforming. 
Lastly, we compute the received power at each test point by adding up the power from each source of any incident ray as received by an isotropic antenna placed at the test point coordinates. 
Note that we assume a power-based association policy, namely we consider each test point to be associated to the \gls{bs} providing the highest power, either over the direct link or via reflections through the deployed \glspl{ris}. 

\begin{figure}[t!]
    \centering  
    \subfigure[\name{} \gls{snr} performance.]
    {
        \includegraphics[clip,width=.46\linewidth ]{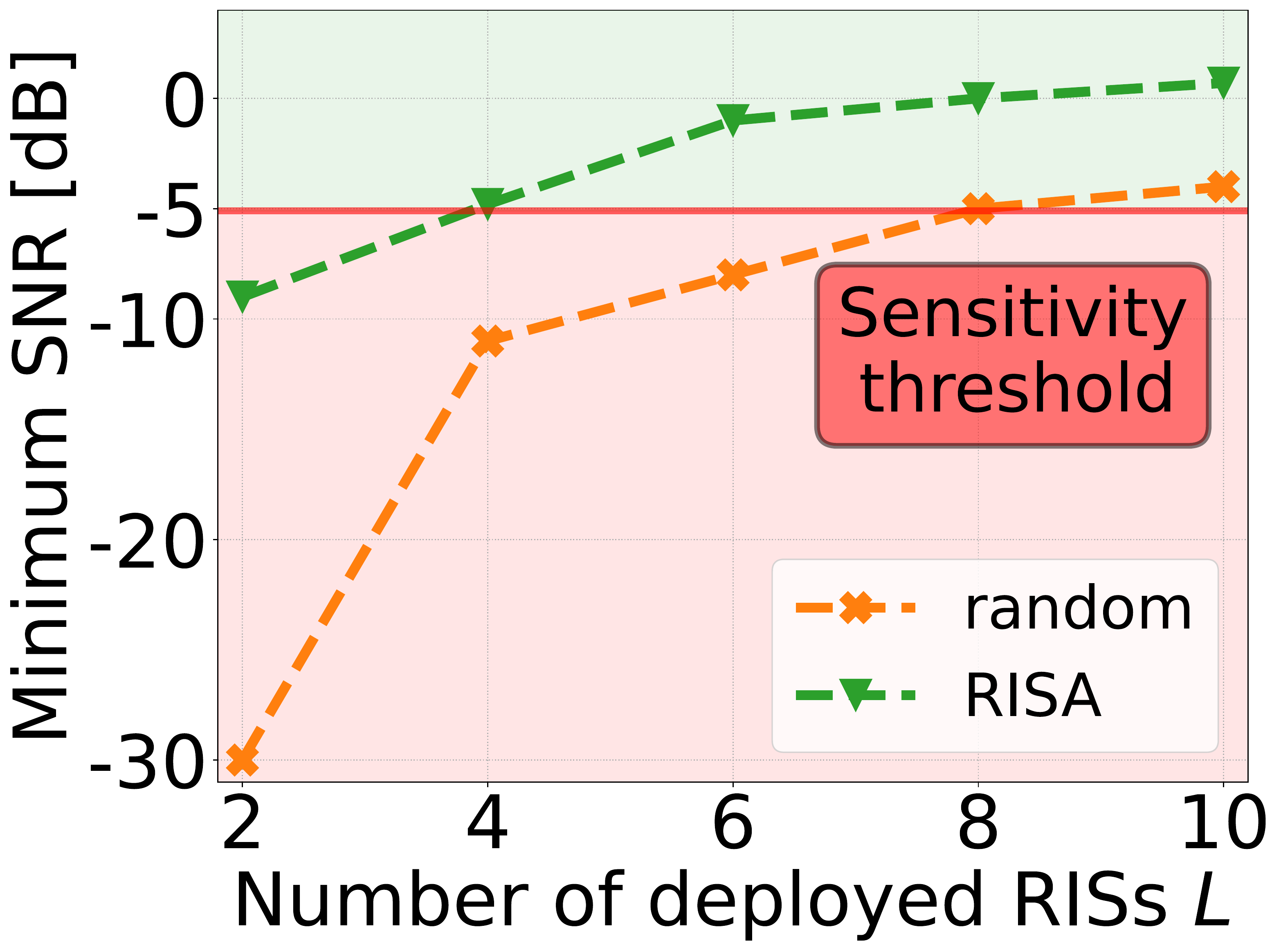}
        \label{fig:snr_vs_nris_rennes}
    }
    \subfigure[\name{} fairness performance.]
    {
        \includegraphics[clip,width=.46\linewidth]{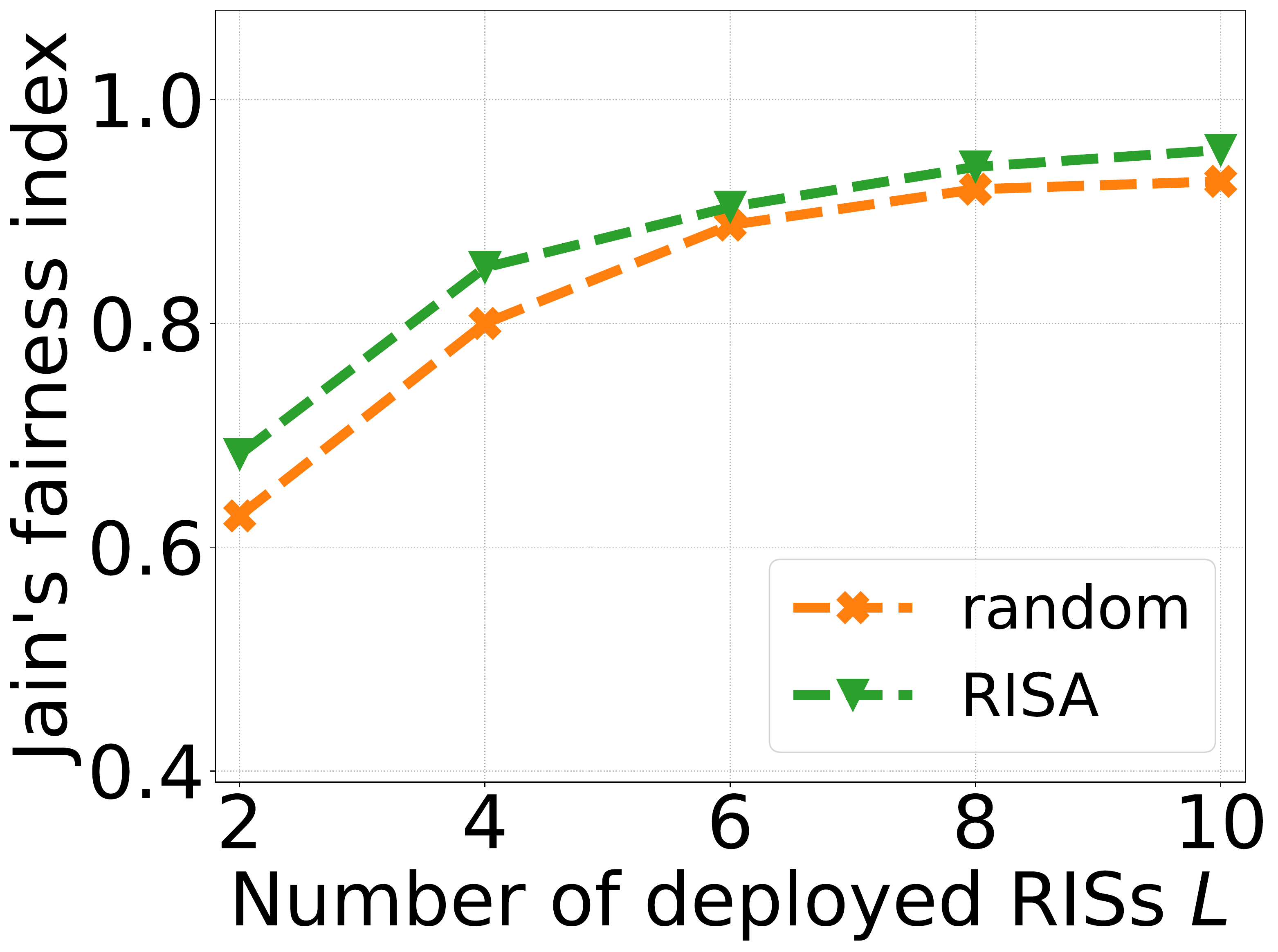}
        \label{fig:jfi_vs_nris_rennes}
    }
    \caption{\label{fig:perf_vs_nris_rennes} \name{} performance obtained via ray-tracing simulations for different numbers of deployed \glspl{ris} and available Candidate Sites (CSs) in a realistic environment (Rennes station).}
\end{figure}

\begin{figure}[t!]
    \centering  
    \subfigure[\gls{snr} heatmap with $L = 0$]
    {
        \includegraphics[clip,width=.46\linewidth ]{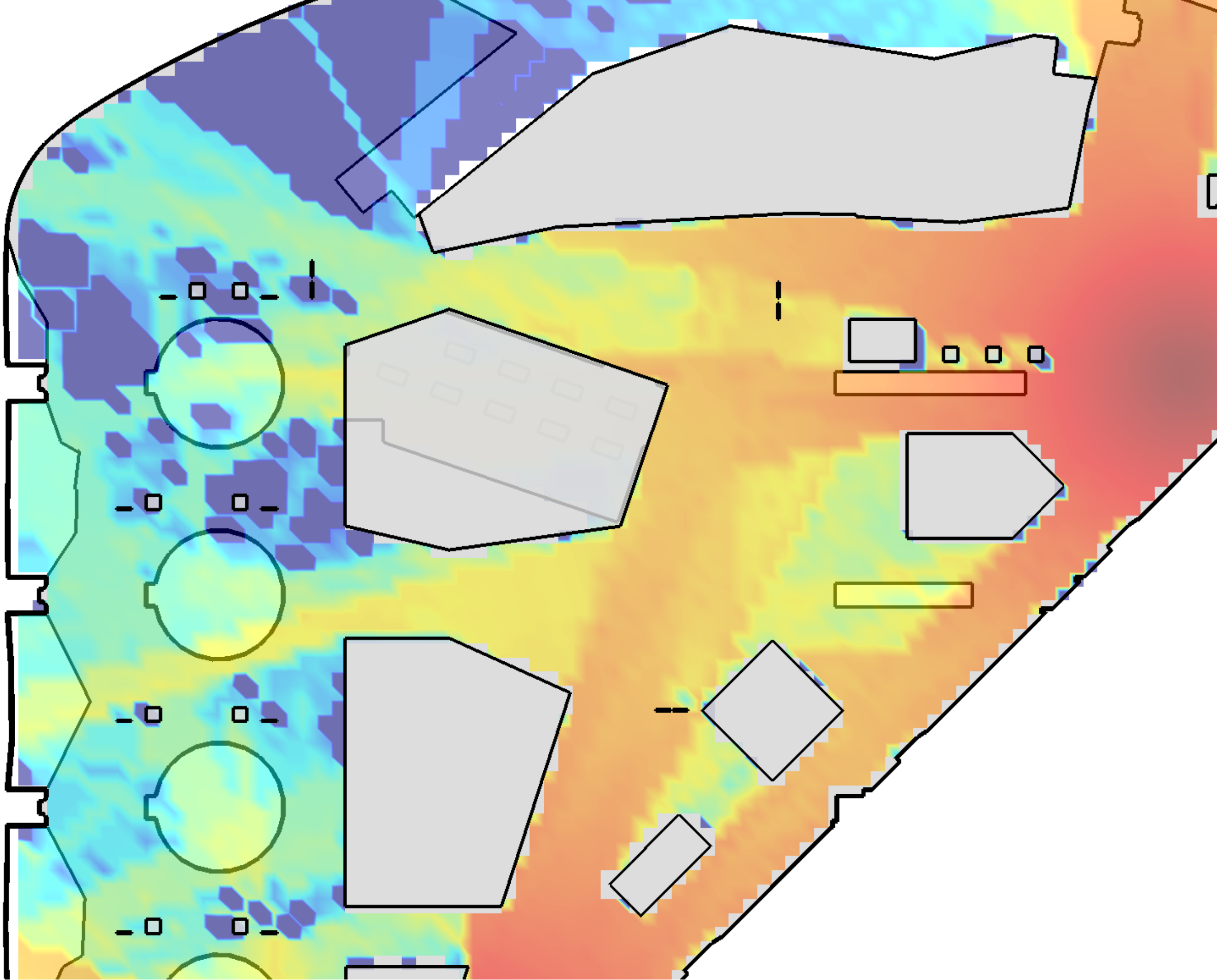}
        \label{fig:2D_heatmap_no_ris}
    }
    \subfigure[\gls{snr} heatmap with $L = 6$ \glspl{ris} (red squares).]
    {
        \includegraphics[clip,width=0.46\linewidth]{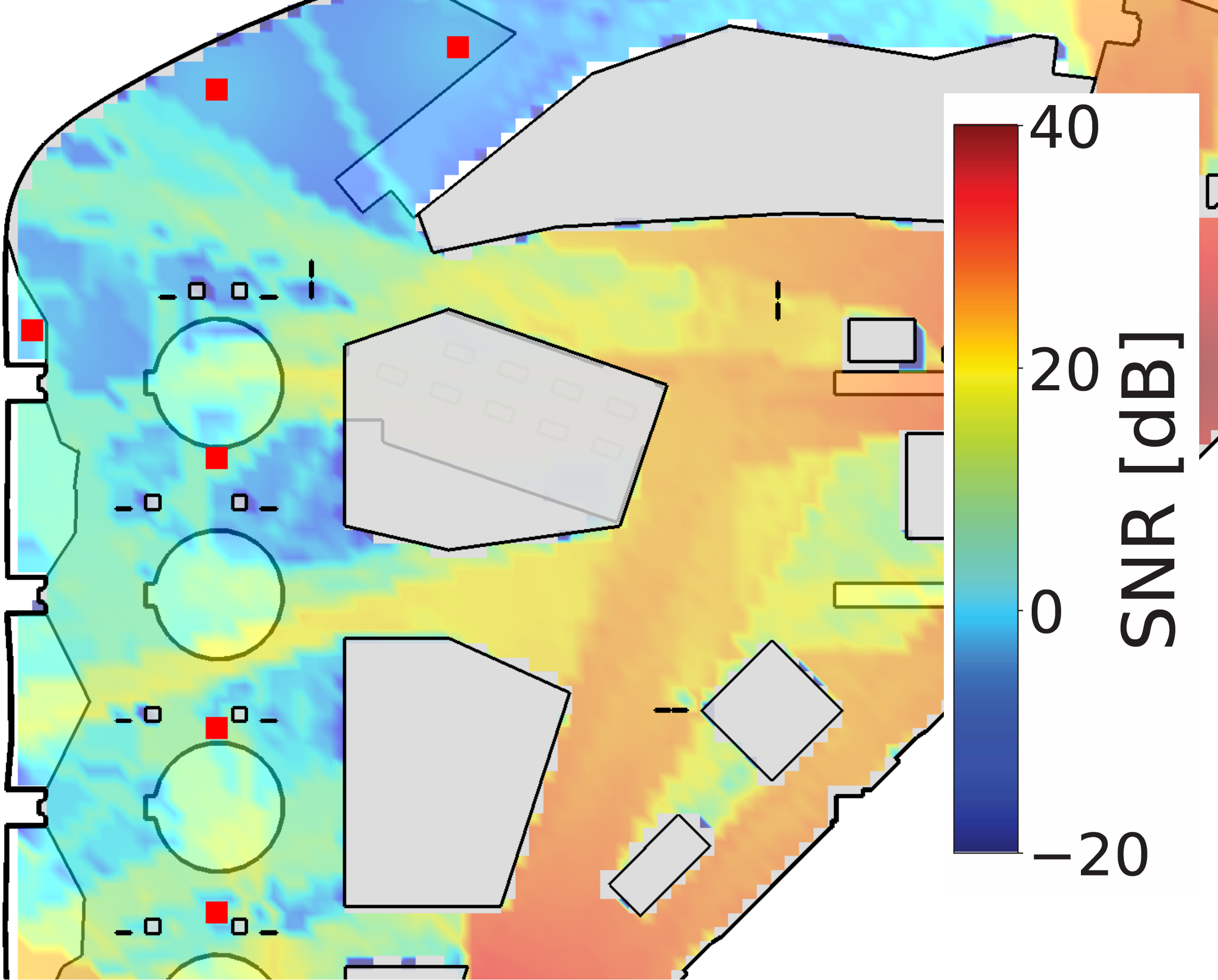}
        \label{fig:2D_heatmap_3_ris}
    }
    \caption{\gls{snr} heatmap in the dead zone (see Figure~\ref{fig:heatmap_no_ris}) of the Rennes station obtained via ray-tracing simulations.}
    \label{fig:heatmap_compaison_rennes}
\end{figure}

\textbf{Realistic simulations.} In Fig.~\ref{fig:perf_vs_nris_rennes}, we show the performance of \name{} for different numbers of deployed \glspl{ris} among $N = 20$ handpicked \glspl{cs} at a height of $5.5$ m and meeting the architectural constraints of the station building. Besides, we compare such results with a random deployment policy averaged over $10^2$ instances. Clearly, \name{} outperforms the random policy in both metrics, i.e., minimum \gls{snr} in Fig.~\ref{fig:snr_vs_nris_rennes} and \gls{jfi} in Fig.~\ref{fig:jfi_vs_nris_rennes}. The fairness is further confirmed in Fig.~\ref{fig:heatmap_compaison_rennes}, wherein we compare the 2D heatmaps of the \gls{snr} obtained by \name{} for $L = 6$ numbers of deployed \glspl{ris} against the baseline with no \gls{ris}.

\section{Conclusions}

\glspl{ris} introduce a novel challenge in traditional cellular networks planning. On the one hand, optimal \glspl{ris} configurations should be computed given fixed \glspl{bs} and \glspl{ris} positions. On the other hand, optimal \glspl{ris} deployments depend on \glspl{ris} configurations. To address these coupled issues and make the analysis tractable, in this paper we proposed \name{}, a RIS-aware network planning solution that builds on double-nested block coordinate ascent to provide an iterative solution to this unprecedented problem. RISA is evaluated on synthetic generic indoor network deployments and in a real railway station (Rennes). Our results show that RISA can $i$) achieve outstanding performance on top of the existing network infrastructure, $ii$) solve the dead-zone problem in highly-crowded environments and $iii$) improve the user fairness at very limited installation costs.



\bibliographystyle{IEEEtran}
\bibliography{IEEEabrv,bibliography}

\begin{thebibliography}{10}
\providecommand{\url}[1]{#1}
\csname url@samestyle\endcsname
\providecommand{\newblock}{\relax}
\providecommand{\bibinfo}[2]{#2}
\providecommand{\BIBentrySTDinterwordspacing}{\spaceskip=0pt\relax}
\providecommand{\BIBentryALTinterwordstretchfactor}{4}
\providecommand{\BIBentryALTinterwordspacing}{\spaceskip=\fontdimen2\font plus
\BIBentryALTinterwordstretchfactor\fontdimen3\font minus
  \fontdimen4\font\relax}
\providecommand{\BIBforeignlanguage}[2]{{%
\expandafter\ifx\csname l@#1\endcsname\relax
\typeout{** WARNING: IEEEtran.bst: No hyphenation pattern has been}%
\typeout{** loaded for the language `#1'. Using the pattern for}%
\typeout{** the default language instead.}%
\else
\language=\csname l@#1\endcsname
\fi
#2}}
\providecommand{\BIBdecl}{\relax}
\BIBdecl

\bibitem{BOL20_ComMag}
E.~Bj\"{o}rnson, O.~\"{O}zdogan, and E.~G. Larsson, ``{Reconfigurable
  Intelligent Surfaces: Three Myths and Two Critical Questions},'' \emph{IEEE
  Communications Magazine}, vol.~58, no.~12, pp. 90--96, 2020.

\bibitem{RIScommag_2021}
E.~C. Strinati, G.~C. Alexandropoulos, H.~Wymeersch, B.~Denis,
  V.~Sciancalepore, R.~D'Errico, A.~Clemente, D.-T. Phan-Huy, E.~De~Carvalho,
  and P.~Popovski, ``{Reconfigurable, Intelligent, and Sustainable Wireless
  Environments for 6G Smart Connectivity},'' \emph{IEEE Communications
  Magazine}, vol.~59, no.~10, pp. 99--105, 2021.

\bibitem{EuCNC_rise6g}
E.~C. Strinati, G.~C. Alexandropoulos, V.~Sciancalepore, M.~Di~Renzo,
  H.~Wymeersch, D.-T. Phan-Huy, M.~Crozzoli, R.~D'Errico, E.~De~Carvalho,
  P.~Popovski, P.~Di~Lorenzo, L.~Bastianelli, M.~Belouar, J.~E. Mascolo,
  G.~Gradoni, S.~Phang, G.~Lerosey, and B.~Denis, ``{Wireless Environment as a
  Service Enabled by Reconfigurable Intelligent Surfaces: The RISE-6G
  Perspective},'' in \emph{2021 Joint European Conference on Networks and
  Communications 6G Summit (EuCNC/6G Summit)}, 2021, pp. 562--567.

\bibitem{albanese_commag21}
A.~Albanese, V.~Sciancalepore, and X.~Costa-Pérez, ``{First Responders Got
  Wings: UAVs to the Rescue of Localization Operations in Beyond 5G Systems},''
  \emph{IEEE Communications Magazine}, vol.~59, no.~11, pp. 28--34, 2021.

\bibitem{DiRenzo2020}
M.~Di~Renzo, A.~Zappone, M.~Debbah, M.-S. Alouini, C.~Yuen, J.~de~Rosny, and
  S.~Tretyakov, ``{Smart Radio Environments Empowered by Reconfigurable
  Intelligent Surfaces: How It Works, State of Research, and The Road Ahead},''
  \emph{IEEE Journal on Selected Areas in Communications}, vol.~38, no.~11, pp.
  2450--2525, 2020.

\bibitem{albanese22}
A.~Albanese, F.~Devoti, V.~Sciancalepore, M.~Di~Renzo, and X.~Costa-P\'erez,
  ``{MARISA: A Self-configuring Metasurfaces Absorption and Reflection Solution
  Towards 6G},'' in \emph{IEEE INFOCOM 2022 - IEEE Conference on Computer
  Communications}, 2022.

\bibitem{di2019smart}
M.~Di~Renzo, M.~Debbah, D.-T. Phan-Huy, A.~Zappone, M.-S. Alouini, C.~Yuen,
  V.~Sciancalepore, G.~C. Alexandropoulos, J.~Hoydis, H.~Gacanin \emph{et~al.},
  ``{Smart radio environments empowered by reconfigurable {AI} meta-surfaces:
  an idea whose time has come},'' \emph{EURASIP Journal on Wireless
  Communications and Networking}, vol. 2019, no.~1, pp. 1--20, May 2019.

\bibitem{metasurface21}
A.~Pitilakis, O.~Tsilipakos, F.~Liu, K.~M. Kossifos, A.~C. Tasolamprou, D.-H.
  Kwon, M.~S. Mirmoosa, D.~Manessis, N.~V. Kantartzis, C.~Liaskos, M.~A.
  Antoniades, J.~Georgiou, C.~M. Soukoulis, M.~Kafesaki, and S.~A. Tretyakov,
  ``{A Multi-Functional Reconfigurable Metasurface: Electromagnetic Design
  Accounting for Fabrication Aspects},'' \emph{IEEE Transactions on Antennas
  and Propagation}, vol.~69, no.~3, pp. 1440--1454, 2021.

\bibitem{Andrews2011}
J.~G. Andrews, F.~Baccelli, and R.~K. Ganti, ``{A Tractable Approach to
  Coverage and Rate in Cellular Networks},'' \emph{IEEE Transactions on
  Communications}, vol.~59, no.~11, pp. 3122--3134, 2011.

\bibitem{Amaldi2003}
E.~Amaldi, A.~Capone, and F.~Malucelli, ``{Planning UMTS base station location:
  optimization models with power control and algorithms},'' \emph{IEEE
  Transactions on Wireless Communications}, vol.~2, no.~5, pp. 939--952, 2003.

\bibitem{Fascista2019}
A.~Fascista, A.~Coluccia, H.~Wymeersch, and G.~Seco-Granados,
  ``{Millimeter-Wave Downlink Positioning With a Single-Antenna Receiver},''
  \emph{IEEE Transactions on Wireless Communications}, vol.~18, no.~9, pp.
  4479--4490, 2019.

\bibitem{papir21}
A.~Albanese, P.~Mursia, V.~Sciancalepore, and X.~Costa-Pérez, ``{PAPIR:
  Practical RIS-aided Localization via Statistical User Information},'' in
  \emph{2021 IEEE 22nd International Workshop on Signal Processing Advances in
  Wireless Communications (SPAWC)}, 2021, pp. 531--535.

\bibitem{Moro2021}
E.~{Moro} \emph{et~al.}, ``{Planning Mm-Wave Access Networks With
  Reconfigurable Intelligent Surfaces},'' in \emph{2021 IEEE 32nd Annual
  International Symposium on Personal, Indoor and Mobile Radio Communications
  (PIMRC)}, 2021, pp. 1401--1407.

\bibitem{Grippo2000}
L.~{Grippo} \emph{et~al.}, ``{On the Convergence of the Block Nonlinear
  Gauss-Seidel Method under Convex Constraints},'' \emph{Oper. Res. Lett.},
  vol.~26, no.~3, p. 127–136, Apr. 2000.

\bibitem{Wu2019}
Q.~Wu and R.~Zhang, ``{Intelligent Reflecting Surface Enhanced Wireless Network
  via Joint Active and Passive Beamforming},'' \emph{IEEE Transactions on
  Wireless Communications}, vol.~18, no.~11, pp. 5394--5409, 2019.

\bibitem{Mursia2021}
P.~Mursia, V.~Sciancalepore, A.~Garcia-Saavedra, L.~Cottatellucci, X.~C.
  Pérez, and D.~Gesbert, ``{RISMA: Reconfigurable Intelligent Surfaces
  Enabling Beamforming for IoT Massive Access},'' \emph{IEEE Journal on
  Selected Areas in Communications}, vol.~39, no.~4, pp. 1072--1085, 2021.

\bibitem{Lu2021}
H.~Lu, Y.~Zeng, S.~Jin, and R.~Zhang, ``{Aerial Intelligent Reflecting Surface:
  Joint Placement and Passive Beamforming Design With 3D Beam Flattening},''
  \emph{IEEE Transactions on Wireless Communications}, vol.~20, no.~7, pp.
  4128--4143, 2021.

\bibitem{Shen2018}
K.~Shen and W.~Yu, ``{Fractional Programming for Communication Systems—Part
  I: Power Control and Beamforming},'' \emph{IEEE Transactions on Signal
  Processing}, vol.~66, no.~10, pp. 2616--2630, 2018.

\bibitem{Sediq2013}
A.~B. Sediq, R.~H. Gohary, R.~Schoenen, and H.~Yanikomeroglu, ``{Optimal
  Tradeoff Between Sum-Rate Efficiency and Jain's Fairness Index in Resource
  Allocation},'' \emph{IEEE Transactions on Wireless Communications}, vol.~12,
  no.~7, pp. 3496--3509, 2013.

\bibitem{Brem2015}
R.~Brem and T.~F. Eibert, ``{A Shooting and Bouncing Ray (SBR) Modeling
  Framework Involving Dielectrics and Perfect Conductors},'' \emph{IEEE
  Transactions on Antennas and Propagation}, vol.~63, no.~8, pp. 3599--3609,
  2015.

\bibitem{rappaport_2001}
T.~S. Rappaport, \emph{Wireless Communications: Principles and Practice, Second
  edition}.\hskip 1em plus 0.5em minus 0.4em\relax Prentice Hall, 2001.

\end{thebibliography}

\end{document}